\documentclass[prl,twocolumn,showpacs]{revtex4}
\usepackage{graphics}
\usepackage{epsfig}
\usepackage{graphicx}

\begin{document}
\title{Valence-Bond-Solid state entanglement in a 2-D Cayley tree}
\author{Heng Fan$^1$, Vladimir  Korepin$^2$, Vwani Roychowdhury$^1$} 
\affiliation{$^1$Electrical Engineering Department,
University of California at Los Angeles, Los Angeles,
CA 90095, USA\\
$^2$C.N.Yang Institute for Theoretical Physics,
State University of New York at Stony Brook, Stony Brook,
NY 11794-3840, USA
}

\date{\today}

\begin{abstract}
The Valence-Bond-Solid (VBS) states are in general ground states for certain
gapped models. We consider the entanglement of 
VBS states on a two-dimensional Cayley tree. 
We show that the entropy of the reduced density operator does not depend
on the whole size of the Cayley tree. We also show that asymptotically, the entropy
is liearly proportional to the number of singlet states
cut by the reduced density operator of the VBS state. 
\end{abstract}

\maketitle

The entanglement of quantum states,
and in particular of the ground states, related with spin systems has been
attracting a great deal of interest, see for example \cite{OAFF,ON,
VLRK,LRV,JK,K,ABV,LO,GDLL,PEDC,HIZ,W,GK,VMC,FKV,DHHLB,VPC,GMC}. 
We can quantify the
entanglement by the von Neumann entropy of the reduced density
operator of the ground states.  
This quantity for discrete spin chain
and lattice models is analogous to the geometric entropy in the
continous field theory\cite{S,CW,HLW}. 
The geometric entropy plays an important role in the quantum field theory and is
considered to be related to the Bekenstein-Hawking black 
hole entropy\cite{BCH}. It has been  suggested\cite{S,CW} that the
geometric entropy is proportional to size of the boundary 
of the block. This problem has been recently studied for 
discrete case in the real free Klein-Gordon fields\cite{PEDC},  
and again, the entropy has been found to be related to the surface area of the
fields.

The entropy of a block of contiguous spins has been obtained
for various one-dimensional spin chains. 
The higher-dimensional spin-chain case is in general more complicated, and only
recently the
entropy of the reduced density operator 
of ground states for higher-dimensional systmes was studied in \cite{PEDC,HIZ,DHHLB,W,GK}.
For discrete 2-D cases, the entropy
is found to be linearly proportional to the boundary size of the reduced density
operator in the lattice\cite{PEDC,HIZ}. The fermion case is also 
studied in \cite{W,GK}.  
In this Letter, we consider the entropy of
a 2-D Valence-Bond Solid (VBS) state on a Cayley tree, which
is the ground state of the Affleck-Kennedy-Lieb-Tasaki (AKLT) model\cite{AKLT,AKLT0}.
The AKLT model is a gapped model \cite{H} and has been well-studied in condensed
matter physics. Certain entanglement properties of the ground state of the
AKLT model were already studied\cite{VMC,FKV,DHHLB,GMC}. But {\em the entropy
of a block of spins in a 2-D Cayley tree case is not available}. 
There are several reasons that make this quantity of interest to the 
condensed matter community. First, the entropy quantifies the entanglement
between the spins in the reduced density operator with the rest of
the ground state. This is important for qunatum computation and
quantum information, and VBS states for quantum computation are 
presnted in Ref.\cite{VC}. 
Secondly, it is interesting to know whether
this quantity is related to some macroscopic properties,  such
as susceptibility as pointed out in Ref.\cite{GRAC}.
Third, it is generally expected that the entanglement inherent in a system
is possibly
responsible for the phenomena of quantum phase-transitions\cite{OAFF}.
Since the AKLT model is a well studied model in condensed matter
physics, it is interesting to
uncover relationships between the entropy of a block of spins in the ground state
and other measurable physical quantities, such as the correlation functions.
Finally, the entanglement properties can indicate  whether the
density matrix renormalization method \cite{W0} can be used to efficiently
simulate the quantum many-body systems\cite{VGC,VPC1}.
 
The Hamiltonian of the AKLT model is written as \cite{AKLT}
\begin{eqnarray}
H=\sum _{(i,j)}P_z(\vec {S}_i+\vec {S}_j)
\end{eqnarray}
where $\vec{S}_j$ is the spin operator on lattice site $i$,
$P(\vec {S})$ is the the orthogonal projection, and
$(i,j)$ are undordered pairs in the lattice.  
The ground states of the AKLT model are known as VBS states, which
are constructed from the singlet state $|\Psi ^-\rangle =\frac {1}{\sqrt{2}}(|\uparrow \downarrow \rangle
-|\downarrow \uparrow \rangle $. 
For convenience sake, we replace
the singlet state by 
$|\Psi \rangle = 
\frac {1}{\sqrt{2}}(|\uparrow \uparrow \rangle
+|\downarrow \downarrow \rangle $. This substituition does not
change any properties of the entanglement of the VBS state in this paper.    

Let's consider a $C_4$ Cayley tree as in Ref.\cite{AKLT} and as 
showed in the Figure 1. 
At each site, three singlet states are conneted to it. 
Each dot represents a spin-$\frac {1}{2}$.
A symmetrization of the three spin-$1/2$'s creates a spin-$3/2$ at each site.
First, we only onsider one lattice site $A$, and three spin $1/2$'s ($A_1, A_2,A_3$) are
located at this site. The three singlet states are $|\Psi \rangle _{A_1B_1}
|\Psi \rangle _{A_2C_2}|\Psi \rangle _{A_3D_3}$; see Fig 2. 

\begin{figure}[ht]
\includegraphics[width=3cm]{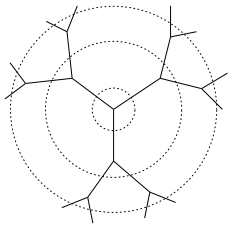}
\caption{Cayley tree: each lattice site has three singlet states conneted
to it, and a symmetrization operator is performed on each lattice site.}
\end{figure}

\begin{figure}[ht]
\includegraphics[width=2cm]{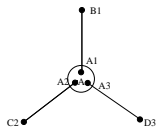}
\caption{One site reduced density operator. 
Site $A$ is four dimension which is a symmetrization of
three spin-1/2 at $A_1,A_2,A_3$. Site $A$ is one part of three singlet states.}
\end{figure}

For the VBS state, we need to consider a symmetrization, $P_A$, at each lattice site $A$.
We realize the symmetrization operator $P_A$ as follows:
\begin{eqnarray}
P_A&=&\frac {1}{4}(|3\uparrow \rangle \langle 3\uparrow |
+|2\uparrow ,\downarrow \rangle \langle 2\uparrow ,\downarrow |
\nonumber \\
&&+|\uparrow ,2\downarrow \rangle \langle \uparrow ,2\downarrow |
+|3\downarrow \rangle \langle 3\downarrow |),
\end{eqnarray}
where state $|i\uparrow ,j\downarrow \rangle $ is a symmetrized state
with $i$ spin up, and $j$ spin down. For example
$|2\uparrow ,\downarrow \rangle =\frac {1}{\sqrt{3}}(
|\uparrow \downarrow \downarrow \rangle
+|\downarrow \uparrow \downarrow \rangle
+|\downarrow \downarrow \uparrow \rangle )$.
For convenience, we also will use the notations
$|3\uparrow \rangle =|4\rangle $,
$|2\uparrow ,\downarrow \rangle =|3\rangle $,
$|\uparrow ,2\downarrow \rangle =|2\rangle $,
$|3\downarrow \rangle =|1\rangle $.
 
{\it One site reduced density operator:}\hspace*{2mm} 
Let's start from lattice site $A$, three singlet states are conneted with site $A$, 
$|\Psi \rangle _{A_1B_1}
|\Psi \rangle _{A_2C_2}|\Psi \rangle _{A_3D_3}$, see Figure 2. If we choose symmetrizatin on $A$, we can
find the following
\begin{eqnarray}
|\Phi \rangle _{A_{123}B_1C_2D_3}&=&P_A
|\Psi \rangle _{A_1B_1}
|\Psi \rangle _{A_2C_2}|\Psi \rangle _{A_3D_3}
\nonumber \\
&=&\frac {1}{2}(|44\rangle +|33\rangle +|22\rangle +|11\rangle ),
\label{state}
\end{eqnarray} 
where one site is lattice site $A$ which is
a symmetrization on spin-1/2 sites $A_1,A_2,A_3$, other sites
are boundary sites $B_1,C_2,D_3$.
We can verify that even if we just project the state at 
lattice site A to the symmetric subspace, then
automatically, the rest of the sites $B_1,C_2,D_3$ are also projected to the corresponding symmetric subspaces.  
The reduced density operator of lattice site $A$ can be found easily:  
$\rho _A=\frac {1}{4}I,$
which is 
the identity operator, $I$, in $SU(4)$ with a normalization factor so that the trace is $1$.
For the Cayley tree case presented in Figure 1, the reduced density operator on each lattice site is
the same except for the boundary sites.
The von Neumann entropy of the reduced density operator is
$S(\rho _A)=2$.
So, we know that {\em for the Cayley tree case, each lattice site is maximally 
entangled with the rest of the sites}, and the entanglement is $2$ ebits.
This result can be directly extended to other types of Cayley tree cases, and we can show  that
each lattice site is maximally entangled with the rest of the lattice 
sites. 
 
{\it The entropy of the reduced density operators does not depend 
on the whole size of the Cayley tree:}\hspace*{2mm} 
Let's study the quantum state of (\ref{state}). 
This is a six-partite
state, and we consider the entanglement across $B_1:A_{1,2,3}C_2D_3$ cut.
It follows directly from the state in (\ref{state}) that the state across 
the cut $B_1:A_{1,2,3}C_2D_3$ cut is maximally entangled. 
So, we can treat the quantum
state $|\Psi \rangle _{B_1:A_{123}C_2D_3}$ in (\ref{state}) as a singlet state
as follows,
\begin{eqnarray}
|\Psi \rangle _{B_1:A_{123}C_2D_3}
=(|\uparrow \Psi _{\uparrow }\rangle _{B_1\bar {B}_1}
+|\downarrow \Psi _{\downarrow }\rangle _{B_1\bar {B}_1})/\sqrt{2},
\label{observe}
\end{eqnarray}
where we denote 
\begin{eqnarray}
|\Psi _{\uparrow }\rangle _{\bar {B}_1}
&=&\frac {1}{\sqrt{2}}|\uparrow \uparrow \rangle _{C_2D_3}|4\rangle _A
+\frac {1}{\sqrt{6}}(|\uparrow \downarrow \rangle _{C_2D_3} 
\nonumber \\
&&+|\downarrow \uparrow \rangle _{C_2D_3})|3\rangle _A
+\frac{1}{\sqrt{6}}|\downarrow \downarrow \rangle _{C_2D_3}|2\rangle _A,
\nonumber \\
|\Psi _{\downarrow }\rangle _{\bar {B}_1}
&=&\frac {1}{\sqrt{6}}|\uparrow \uparrow \rangle _{C_2D_3}|3\rangle _A
+\frac {1}{\sqrt{6}}(|\uparrow \downarrow \rangle _{C_2D_3} 
\nonumber \\
&&+|\downarrow \uparrow \rangle _{C_2D_3})|2\rangle _A
+\frac {1}{\sqrt{2}}|\downarrow \downarrow \rangle _{C_2D_3}|1\rangle _A.
\label{Bbar}
\end{eqnarray}
This can also be understood as the following: First, we have a
singlet state shared by $B_1A_1$. 
To expand the Cayley tree from $A_1$ to $A_{123}C_{23}$ (i.e.,
one leg is expanded to two legs), 
we put two additional singlet
states shared by $A_2,C_2$ and $A_3,C_3$, and perform the symmetrical
projection on site $A_{123}$. 
The final result is  that we just replace the spin up and spin down in $A_1$
by $|\Psi _{\uparrow }\rangle $ and $|\Psi _{\downarrow }\rangle $. 
Similarly, we can further expand the Cayley tree from sites $C_2$, and $C_3$.
If we only consider the {\it bipartite} entanglement of the VBS state 
on a Cayley tree, the bipartite state $B_1{\bar {B}}_1$ is just
a singlet state no matter how many legs are represented by $\bar {B}_1$.
  
In this Letter, we will consider the entropy of reduced density operators
of the VBS on a Cayley tree. The reduced density operators are
one-site, 4-site,..., $\sum _N3\times 2^N+1$-site in size. In Figure 1,
the cuts are shown as dashed circles. The entropy of the reduced density operator
is the {\it bipartite} entanglement across these cuts, i.e., between the spins in the 
reduced density operator and the rest. 

Based on our observations, we present one of our main conclusions in this Letter: 
{\it The entropy of reduced density operator
$S(\rho _i)$ does not depend on the whole size of the
Cayley tree, where $i=1,4,...$}. 
The reason follows from our observation in Eq.(\ref{observe}), which implies that
for the reduced density operator $\rho _i$, the legs connected to it in the
Cayley tree can be dealt as just singlet states, no matter how large the
Cayley tree itself is. We can repeatedly
use this relation and expand the Cayley tree, but the quantity $S(\rho _i)$ does not change.  
So, to study the quantity $S(\rho _i)$, we only need to consider the 
smallest Cayley tree since the entropy does not
denpend on the size of the Cayley tree. 
For example, {\em if we consider $S(\rho _4)$, we 
will just need to study the six legs case as in Figure 3}.
This result is similar to the result for 1-D case\cite{FKV} where
a different proof is given. We should note that the restriction
on $i$ can be relaxed, even though we only consider special cases in
this Letter.

\begin{figure}[ht]
\includegraphics[width=2cm]{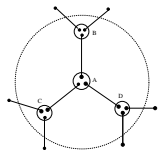}
\caption{4 sites Cayley tree. To study the entropy of
4-site reduced density operator, we can choose the smallest
Cayley tree. There are 6 legs (singlet states) 
connected to the 4-site
reduced density operator}
\end{figure}

\begin{figure}[ht]
\includegraphics[width=1cm]{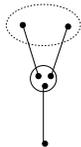}
\caption{Entropy of reduced density operator of two legs (dashed circle) is $\log 3$}
\end{figure}

{\it Entropy of 4-site reduced density operator:}\hspace*{2mm}
For a biparite pure state $|\Psi _{AB}\rangle $, we know
$S(\rho _A)=S(\rho _B)$, where $\rho _{A(B)}$ is the reduced density
operator. To calculate the entropy of the 4-site reduced
density operator, we can calculate the entropy of
the reduced density operator of six legs; see Figure 3.
First we have the state 
$|\Psi _{A,B_1C_2D_3}\rangle $ which takes the form (\ref{state})
as in Figure 2. Then we
can substitute the spin up and spin down in $B_1,C_2,D_3$ by 
$|\Psi _{\uparrow }\rangle $ and $|\Psi _{\downarrow }\rangle $, and 
now the correponding Cayley tree reduces to that shown in Figure 3.
After
tracing over spin-$2/3$ sites in $ABCD$, we obtain the reduced density
operator as follows:
\begin{eqnarray}
\bar {\rho }_4=
\frac {2}{3}I^{\otimes 3}+\frac {1}{6}
\rho _{\uparrow }^{\otimes 3} 
+\frac {1}{6}
\rho _{\downarrow }^{\otimes 3}
+X,
\label{reduce4}
\end{eqnarray}
where $I=\frac {1}{2}(\rho _{\uparrow }+\rho _{\downarrow })=diag. (1/3,1/3,1/3)$, 
$\rho _{\uparrow }=diag.(1/2,1/3,1/6)$,
$\rho _{\downarrow }=diag.(1/6,1/3,1/2)$,
the basis chosen here is $\{ |2\uparrow \rangle ,|\uparrow ,\downarrow \rangle ,
|2\downarrow \rangle \} $ which are basis vectors  of the symmetrical subspace.
Moreover, $\rho _{\uparrow }(\rho _{\downarrow })$ is derived from 
$|\Psi _{\uparrow }\rangle (|\Psi _{\downarrow }\rangle )$ by taking
the trace on the spin $2/3$ site,
\begin{eqnarray}
X&=&\frac {1}{108}[I\otimes (x\otimes x^t+x^t\otimes x)
+(x\otimes x^t+x^t\otimes x)\otimes I
\nonumber \\
&&+x\otimes I\otimes x^t
+x^t\otimes I\otimes x],
\label{offdiag}
\end{eqnarray} 
where $x=|2\uparrow \rangle \langle \uparrow ,\downarrow|
+|\uparrow ,\downarrow \rangle \langle 2\downarrow |$.
Since the entropy of the density operator of $ABCD$
is equal to six legs conneted to it, $S(\rho _4)=S(\bar {\rho _4})$,
the entropy of the 4-site reduced density operator
is
\begin{eqnarray}
S(\rho _4)=\frac {53}{54}+\frac {13}{4}\log 3-\frac {65}{108}\log 5
\approx 4.735,
\end{eqnarray}
where $\log $ has the base $2$. 
We can find that $\rho _4$ is actually like $I^{\otimes 3}$ which
has the entropy $3\log 3\approx 4.755$. The difference between these two
entropies is less than $0.5\% $. In fact, the off-diagonal entries
$X$ in (\ref{reduce4}) are small, and 
$\frac {1}{2}\sum _{\alpha =\uparrow \downarrow }\rho _{\alpha }^{\otimes 3}$
is close to $I^{\otimes 3}$. For simplicity, we can use 
the fidelity to define the
distance between $\rho _{\uparrow ,\downarrow }$ and $I$. The fidelity
is defined as $F(\rho _A,\rho _B)=Tr\sqrt{\sqrt{\rho _A }\rho _B\sqrt{\rho _A}}$.
The fidelity is $0.977$ for both $\rho _{\uparrow }$ and $\rho _{\downarrow }$, 
and we know that $\rho _{\uparrow ,\downarrow }$
is like the identity $I$. That's the reason why $S(\rho _4)\approx I^{\otimes 3}$.  

{\it General entropy of reduced density operator of the VBS state in Cayley tree:} \hspace*{2mm}
In the 2-D case presented in Ref.\cite{PEDC,HIZ}, the entropy of the reduced
density operator is linearly proportional to size of the boundary. 
A toy model can be considered
for VBS states: For the AKLT model, the ground states are constructed from the
singlet states by some projection operators. Suppose we consider a state without
any projections, the entropy of this state is the number of singlet states that cross
the boundary. The result that the entropy is linearly proportional to 
the boundary size can be roughly understood as that the entropy is linearly proportional
to the number of states (like the singlet states) cut by the boundary.  
For the VBS state in a Cayley tree, we expect similar results.
In the last section, we found that $S(\rho _4)\approx  3\log 3$. The number of singlet
states crossing the boundary is 6. We can expect that in general we have
$S(\rho _i)\approx \frac {N}{2}\log 3$, i.e., $N$ singlet states cross the 
boundary ($i=\sum _k3\times 2^k+1$, here $N$ and $i$ has a restriction between them).

Let's show next the result: {\it The
entropy of the reduced density operator $S(\rho _i)$ is upper bounded by $\frac {N}{2}\log 3$}.
A simple observation is that the upper bound is $N$ since $N$ singlet states cross the boundary
and the projection can only reduce this quantity.
We next show that a tighter upper bound can be found. Consider the state presented in (\ref{state}),
the reduced density  operator crossing two singlet states is $diag.(1/3,1/3,1/3)$,
and the entropy is $\log 3$, see Figure 4. 
This can be considered as a building block for the Cayley tree,
thus we know that $S(\rho _i)\le \frac {N}{2}\log 3$.


This is just the first order upper bound, and we can further consider a second order 
upper bound: To find the entropy of 4 legs, for this case, we suppose the Cayley tree
is relatively large. Before we present this second order upper bound, we present the
techniques used in our calculations by the following example. 

We use $\rho _{10}$ as an example. 
Starting from $\rho _4$ in (\ref{reduce4}), we can expand the Cayley tree by 
substituting the state $|\uparrow \rangle $ and $|\downarrow \rangle $ by
$|\Psi _{\uparrow }\rangle $ and $|\Psi _{\downarrow }\rangle $, respectively.
By tracing out the spin-$2/3$ sites, we can obtain $\rho _{10}$
from $\rho _4$ in (\ref{reduce4}) by the following substitutions,
\begin{eqnarray}
\rho _{\alpha }&\rightarrow &\frac {2}{3}I^{\otimes 2}+\frac {1}{3}
\rho _{\alpha }^{\otimes 2}+Y, ~~\alpha =\uparrow ,\downarrow,
\label{abexpand}\\
x&\rightarrow &\frac {1}{12}(x\otimes I+I\otimes x),
\label{xexpand}\\
Y&=&\frac {1}{108}(x\otimes x^t+x^t\otimes x).
\label{expand}
\end{eqnarray}
The final result shows that $\rho _{10}\approx I^{\otimes 6}$, 
and thus $S(\rho _{10})\approx 6\log 3$. Actually, 
we find $S(\rho _{10})\approx 9.4891\approx 6\log 3(1-\epsilon )$,
and $\epsilon $ is small and is about $0.22\%$. 
The fidelity between 
the term related with $\rho _{\alpha }$ in $\rho _{10}$ 
with identity is
0.994 which is better than the original 0.977 in $\rho _4$ case. 
The Eqs.(\ref{abexpand},\ref{xexpand},\ref{expand})
and Eq.(\ref{reduce4}) provide us with an algorithm to find the
general reduced density operators of the VBS state on a Cayley tree.

For the general case, we can expand the Cayley tree and obtain the reduced density operator $\rho _{i}$
by relations (\ref{abexpand},\ref{xexpand},\ref{expand}).
Suppose $i=\sum _{k=0}^M 3\times 2^k+1$; then 
the general form will be as follows: 
\begin{eqnarray}
\rho _i&=&\frac {2}{3}I^{\otimes 3\times 2^{M-1}}
+\frac {1}{6}\sum _{\alpha =\uparrow ,\downarrow }(f(\rho _{\alpha },I))
+g(x,I)\nonumber \\
&\approx &I^{\otimes 3\times 2^{M-1}}.
\end{eqnarray}
The last equation follows from the fact that the second term is more like the 
identity operator as $i$ becomes large, and the
third term remains small and can be omitted. 
We know that $S(\rho _i)\approx 3\times 2^{M-1}\log 3, M\ge 1$. The reduced density operator
cuts $N=3\times 2^M$ singlet states. And we thus estimate that {\it asymptotically the entropy of VBS
in a Cayley tree is linearly proportional to the number of singlet states across the
boundary}. 

Wee should point out that though $S(\rho _i)\approx 3\times 2^{M-1}\log 3, M\ge 1$,
the upper bound $3\times 2^{M-1}\log 3$ can never be saturated. The reason is that, as we pointed
out, this is the first order upper bound. We can also compute a second order upper bound.
Using the same method as presented above, we can obtain the reduced density opertor
of 4 legs: Start from the 2-leg reduced density operator
$|2\uparrow \rangle \langle 2\uparrow |+
|\uparrow ,\downarrow \rangle \langle \uparrow ,\downarrow |
+|2\downarrow \rangle \langle 2\downarrow |$, replace the state $|\uparrow \rangle $
by $|\Psi _{\uparrow }\rangle $ and similarly for $|\downarrow \rangle $, trace out
the corresponding sites, and then we can find the entropy of 4 legs to be 
$S_4
\approx 3.1631$.
We know the first order upper bound should give $2\log 3\approx 3.1699$. The difference between these two
bounds is about $0.21$ percent. 
So we have a more exact result: $S(\rho _i)\le 3\times 2^{M-2}\times 3.1631$.
Of course, we can also compute higher-order upper bounds. But as in 
1-D VBS state in \cite{FKV}, the correction of higher order bounds will decay exponentially.
We performed the calculations for the density operator of 8 legs and 16 legs
corresponding to third and fourth order corrections for the upper bound.
As expected, we get $S_8/4\log 3=(1-\epsilon ')$ and
$S_{16}/8\log 3=(1-\epsilon '')$, both $\epsilon '$ and $\epsilon ''$ are around
$0.21\% $ which have almost no difference with the second order correction.
So roughly, the first order upper bound $\log 3$ still works.

Observing that the result of $S(\rho _{10})$ actually provides 
a lower-bound correction for
the general $S(\rho _i)$, when $i$ is large, asympototically we have
\begin{eqnarray}
1-\epsilon _l\le S(\rho _i)/3\times 2^{M-1}\log 3
\le 1-\epsilon _u
\label{result}
\end{eqnarray}
where safely we can set 
$\epsilon _l=0.23\% $,  and $\epsilon _u=0.20\% $.
Now we summarize our main result: {\it Asymptotically, the entropy of the
VBS state in a Cayley tree is linearly proportially to the number of
singlet states that cross the boundary.} 
We estimate that asymptotically $\epsilon \approx 0.22\%$.  

We estimated the entropy of the reduced density operators that are circles, 
as presented in Figure 1. But our method works for all kinds of reduced
density operators on the Cayley tree. 
The only difference is that the substitutions 
(\ref{abexpand},\ref{xexpand}) 
will depend on the form of the reduced density operators
in the Cayley tree. 
We expect that our result
that asymptotically, the entropy is liearly proportinal to the number of singlet states
across the boundary, still holds given that the number of cut singlet
states is large. We performed the calculations for a density
operator with 16 legs (not a circle) extended from $\rho _{10}$, and we found that 
the Eq. (\ref{result}) still holds for this case. Some other forms, with fewer
legs,  have been also checked and our conjecture holds for all these cases as well. 

In fact, we actually provide an algorithm to find the 
reduced density operators
of VBS state on a Cayley tree. It will be interesting
to apply this algorithm in the simulations of quantum
many-body systems.

\noindent {\it Acknowlegements}: H.F. and V.R. are
supported in part by the U.S. Army Research 
Office/DARPA under contract/grant number DAAD 19-00-1-0172,
and in part by the NSF under contract number CCF-0432296.
We would like to thank Frank Verstraete for useful discussions.

\end{document}